\documentclass{aa}  
\usepackage{graphicx}
\usepackage{txfonts}

\usepackage{hyperref}
\hypersetup{
    colorlinks=true,
    linkcolor=cyan,
    citecolor=cyan,
    urlcolor=cyan,
    }

\begin{document}

    
    \title{Unequivocal detection of the tidal deformation of a red giant in~a~binary system via interferometry}
    \titlerunning{Tidal deformation of HD 352}
    
   \author{Jaroslav Merc\inst{1, 2, 3}
          \and
          Henri\,M.\,J. Boffin\inst{1}
                    }

\institute{European Southern Observatory, Karl-Schwarzschild-Stra\ss{}e 2, Garching bei M\"{u}nchen, 85748, Germany
\and
Astronomical Institute of Charles University, V Hole\v{s}ovi\v{c}k{\'a}ch 2, Prague, 18000, Czech Republic
\and
             Instituto de Astrof\'isica de Canarias, Calle Vía Láctea, s/n, E-38205 La Laguna, Tenerife, Spain\\
    \email{jaroslav.merc@gmail.com}
    }

   \date{Received August 22, 2024; accepted November 21, 2024}


  \abstract
   {While mass transfer in binary systems is a crucial aspect of binary evolution models, it remains far from understood. 
HD~352 is a spectroscopic binary exhibiting ellipsoidal variability, likely due to a tidally deformed giant donor filling its Roche lobe and transferring matter to a faint companion. Here, we analyse VLTI/PIONIER interferometric observations of the system, obtained between 2010 to 2020. We demonstrate that observations near the system's quadrature cannot be explained by simple symmetric disk models, but are consistent with the shape of a Roche-lobe-filling star. We think that this is the first case of tidal deformation of a red giant being observed directly, thanks to the interferometric technique. By combining our interferometric modeling results with the analysis of the optical spectrum, multi-frequency spectral energy distribution, and published radial velocities and light curves, we constrain the system parameters and show that HD~352 will likely soon enter the common envelope phase, although we cannot reject the hypothesis that it is undergoing stable mass transfer against theoretical predictions. This has important consequences for modeling a large class of binary systems. Additionally, our observations confirm that Roche-lobe-filling giants can be resolved with interferometry under favorable conditions. Such observations may help resolve the mass transfer dichotomy in systems like symbiotic binaries, where the predominant mass transfer mode remains unclear.}


   \maketitle
%

\section{The interacting star HD 352} \label{sec:intro}

Most stars are born with a companion \citep{2017ApJS..230...15M,2023ASPC..534..275O} and in many such systems, the components undergo interactions at various stages of their evolution. However, numerous open questions remain regarding binary interactions, particularly those related to mass transfer and accretion \citep{2023hxga.book..129B}. Addressing these questions is crucial, as such interacting binaries lead to some of the most interesting stellar phenomena, such as Type Ia supernovae, luminous red novae, or gravitational wave emitters.

The binarity of HD~352 (AP Psc; sometimes denoted as 5 Ceti, but actually located in Pisces) was first reported in 1914 based on radial velocity variations \citep{1914PASP...26..261A}. The first spectroscopic orbit, presented by \citet{1933ApJ....77..310C}, suggested an orbital period of slightly above 96 days and an eccentricity of 0.12. Since then, the binary received little attention, except for revisions of the eccentricity to near zero \citep{1971AJ.....76..544L,1985PASP...97..355B}, until the discovery of photometric variability in 1981 \citep{1981IBVS.2013....1L}. These authors proposed that the observed variability was due to eclipses similar to W UMa-type binaries, even if they were puzzled by its occurrence in a long-period system. This contact binary model led some authors to propose an enormous total system mass exceeding 100 M$_\odot$ \citep{1989ChA&A..13..181L}. In contrast, \citet{1986IBVS.2952....1E,1988AcA....38..353E} suggested that the variability was due to the ellipsoidal effect in a semi-detached binary consisting of a K giant and a companion resembling a mid F-type main-sequence star, but with higher luminosity due to accretion. They estimated the masses of both components to be approximately 1-2 M$_\odot$. 
Revised orbital elements were published by \citet{2007PASP..119..886E} and \citet{2011MNRAS.410.1761K}. Both studies reported a slight eccentricity of the orbit, e $\sim$ 0.02-0.03. Interestingly, subsequent studies by both groups argued that this eccentricity is caused by non-Keplerian velocity perturbations induced by the distortion of the giant star filling its Roche lobe \citep{2008ApJ...681..562E,2012MNRAS.427..298H}.

\citet{2014A&A...564A...1B} analyzed a single interferometric observation of HD~352 obtained using the PIONIER instrument on the ESO's Very Large Telescope Interferometer (VLTI) and concluded that the giant was likely elongated. Both elongated Gaussian and elongated disk models provided significantly better fits compared to the uniform symmetric disk model. In this study, we extend on that work and aim to investigate multiple VLTI/PIONIER observations of HD~352, covering various orbital phases of the system. We show that the giant in HD~352 is indeed tidally distorted.

\section{Methods}\label{sec:methods}

\begin{table}[h]
\centering
\caption{Log of VLTI/PIONIER observations.}\label{tab:log}%

\begin{tabular}{@{}llc@{}}
\hline
\noalign{\smallskip}
Night & Config.  & Orbital phase\\
\hline
\noalign{\smallskip}
04 December 2010    & E0-G0-H0-I1 &  0.79\\
13 August 2012    & A1-G1-I1-K0 & 0.20\\
15 October 2013    & A1-G1-J3-K0 & 0.64\\
22 September 2014    & A1-G1-J3-K0 &  0.19\\
29 July 2019    & A0-B2-C1-D0 &  0.55\\
22 December 2019    & D0-G2-J3-K0 &  0.06\\
12 December 2020    & A0-G1-J2-J3 & 0.76\\
18 December 2020*    & D0-G2-J3-K0 &  0.82\\
19 December 2020*    & D0-G2-J3-K0  &  0.83\\
27 December 2020    & A0-B2-C1-D0 &  0.91\\
\hline
\end{tabular}
\tablefoot{*These two observations were analyzed together.}

\end{table}

\subsection{Observations and data reduction}
HD~352 was observed in the $H$-band with the four 1.8-m Auxiliary Telescopes of VLTI using the PIONIER instrument \citep{2011A&A...535A..67L} on 10 individual nights over a 10-year period, from 2010 to 2020. The log of the observations is shown in Table~\ref{tab:log}, which indicates that the data were taken with different baseline configurations, some small and others large. Observations on 29 July 2019 and 27 December 2020 used the small configuration with short baselines, which did not provide sufficient angular resolution for meaningful analysis (as they probe only scales of a few milli-arcseconds -- mas), and were therefore excluded from further analysis. Moreover, datasets from two consecutive nights in December 2020 were analyzed together. The uv-plane coverage, squared visibilities, and closure phases of all the observations are shown in the appendix (Figs. \ref{fig:interf_2010} - \ref{fig:interf_2020_3} and Figs. \ref{fig:interf_2010_CP} - \ref{fig:interf_2020_3_CP}).
Data reduction was performed using the standard {\tt pndrs} package \citep{2011A&A...535A..67L}, except for observations from 2010 for which the raw data are not available in the ESO Archive. For that epoch, we downloaded the calibrated files from the Jean-Marie Mariotti Center Optical Interferometry DataBase\footnote{http://oidb.jmmc.fr}.

The archival optical spectrum was obtained using the Fiber-fed Extended Range Optical Spectrograph (FEROS) on the MPG/ESO 2.2-m telescope at La Silla, Chile, on 11 July 2003. Three individual exposures of 100 seconds were reduced using the MIDAS data reduction software, and a median spectrum was produced to increase the signal-to-noise ratio. The resolution of the spectrum is R $\sim$ 48\,000, covering the wavelength range of $\sim$350 to 920\,nm.


\begin{figure}[]
\centering
\includegraphics[width=\columnwidth]{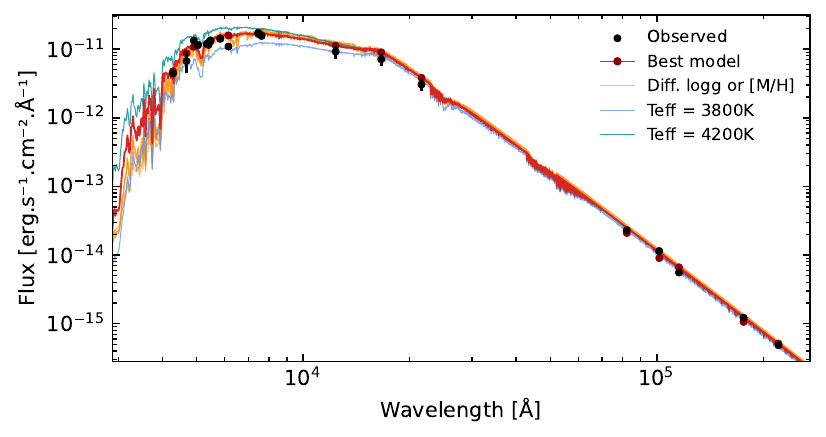}
\caption{Spectral energy distribution of HD~352 (black), along with the best-fitting BT-Settl theoretical spectrum (red spectrum with dark red symbols for calculated fluxes; T$_{\rm eff}$ = 4\,000 K, $\log g$ = 1.0, [M/H] = $-1.0$) obtained using VO SED Analyzer. Additional theoretical spectra with the same T$_{\rm eff}$ but different $\log g$ or [M/H] are shown in orange. Theoretical spectra for T$_{\rm eff}$ = 3\,800 K and 4\,200 K are shown in green and blue, respectively.}\label{fig:sed}
\end{figure}

\subsection{Spectral energy distribution of HD~352}
The spectral energy distribution (SED) of HD~352 (Fig.~\ref{fig:sed}) was constructed on the basis of the multi-frequency photometric observations from the Tycho-2 catalogue \citep{2000A&A...357..367H}, AAVSO Photometric All-Sky Survey \citep[APASS; ][]{2014CoSka..43..518H}, catalog of homogeneous means in the UBV system \citep{2006yCat.2168....0M}, Hipparcos satellite \citep{1997ESASP1200.....E}, \textit{Gaia} DR3 \citep{2016A&A...595A...1G,2023A&A...674A...1G}, Two Micron All Sky Survey \citep[2MASS; ][]{2006AJ....131.1163S}, AKARI satellite \citep{2010A&A...514A...1I}, Infrared Astronomical Satellite \citep[IRAS; ][]{1984ApJ...278L...1N}, and Wide-field Infrared Survey Explorer \citep[WISE; ][]{2010AJ....140.1868W}.

The SED was fitted with the BT-Settl grid of theoretical spectra \citep{2014IAUS..299..271A} using the VO SED Analyzer (VOSA)\footnote{http://svo2.cab.inta-csic.es/theory/vosa/} on the Spanish Virtual Observatory theoretical services website \citep{2008A&A...492..277B}. A distance of 361.3\,$\pm$\,8.8 pc \citep{2023A&A...674A...1G} was adopted, and a range of extinctions A$_{\rm V}$ = 0.00 -- 0.13 mag was allowed \citep[the total Galactic extinction in the direction of HD~352 is about A$_{\rm V}$ = 0.1 mag, according to the dust map by][]{2011ApJ...737..103S}.

\begin{figure}[]
\centering
\includegraphics[width=0.49\textwidth]{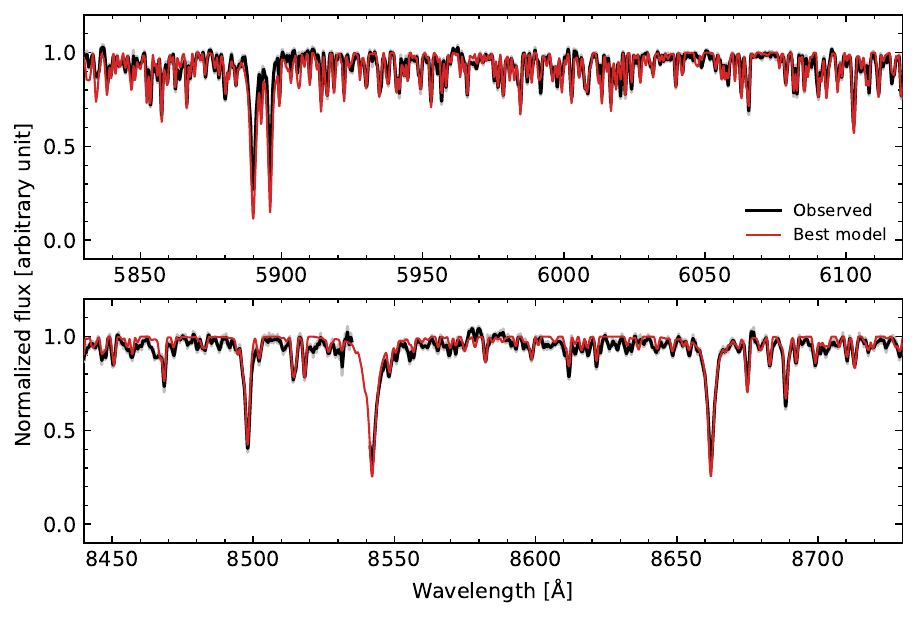}
\caption{Parts of the analyzed median FEROS spectrum (original in gray, smoothed in black), along with the best-fitting spectrum obtained with {\tt iSpec} (red; fixed T$_{\rm eff}$ = 4\,000 K, $\log g$ = 1.27, [M/H] = -0.13).}\label{fig:spectrum}
\end{figure}

\subsection{FEROS high resolution spectrum}
We analyzed the FEROS spectrum (Fig.~\ref{fig:spectrum}) using {\tt iSpec} \citep{2014A&A...569A.111B}. The spectrum was smoothed by convolving it with a Gaussian -- we verified that this does not affect the results. The value of the projected rotational velocity, $v\sin i$,  was obtained from the spectrum using the FEROS cross-correlation function calibration by \citet{2001A&A...375..851M}. The spectrum was then continuum normalized, and stellar parameters were determined through spectral synthesis in {\tt iSpec} using the {\tt MOOG} code \citep{2012ascl.soft02009S}, MARCS model atmospheres \citep{2008A&A...486..951G}, and the Gaia-ESO Survey line list (version 5.0) with hyperfine structure and isotopes \citep{2021A&A...645A.106H}. Solar abundances were taken from \citet{2009ARA&A..47..481A}. The effective temperature, $v\sin i$, and resolution were fixed during the analysis. Only selected spectral regions were considered during the $\chi^2$ calculation ({\tt iSpec}-recommended lines in the optical range, with an additional spectral segment in the near-infrared).

\subsection{Interferometric model}\label{sec:methods_model}

As a first approach, we attempted to fit the interferometric observables using the LITpro model fitting software \citep{2008SPIE.7013E..1JT}. Simple models such as a uniform disk, an elongated disk, a Gaussian, or an elongated Gaussian did not fit the observed data well, particularly for observations obtained with longer baselines and near the quadrature of the system. We note that image reconstruction is unsuitable with our dataset as it requires observations obtained with different configurations over a few days only - at most a few weeks. Indeed, given the relatively short orbital period, our view of the system changes quickly, making it impossible to combine observations on longer timescales.

In the initial LITpro analysis, the combination of two disks (one smaller and shifted to the edge of the other) or a disk and a nearby point source provided visibilities similar to the observed data. Given that the companion is likely too faint in the $H$-band and should be located further from the primary’s disk than this model implies, this additional signal was unlikely due to the companion. Instead, as we can assume that the giant is filling its Roche lobe (based on the ellipsoidal variability and the RV curve), this preliminary observation motivated us to build a model of a Roche-lobe filling star. Such a model is not available in standard codes \citep[e.g., LITpro, OITOOLS, or PMOIRED;][]{2022SPIE12183E..1NM}. However, OITOOLS, written in {\tt Julia}, allows the computation of interferometric observables for any user-provided image model.

We built an input model image of the Roche-lobe filling star in {\tt Python}, assuming the star fully fills its Roche lobe and its surface follows the critical gravitational equipotential, calculated using the Python package {\tt PyAstronomy}, with the apex at the L1 Lagrangian point. The mass ratio was iteratively adopted based on initial calculations from the orbital inclination we obtained from our fit to the visibilities and the known mass function, resulting in a final model with a mass ratio of the giant to its hotter companion, $q = M_{\rm G}/M_{\rm h} = 1.7$. The flux of the star was not assumed uniform; instead, we adopted the power-2 limb-darkening law of \citet{2023A&A...674A..63C}. The coefficients were adopted for a star with T$_{\rm eff}$ = 4\,000 K and $\log g$ = 1.3 dex, in accordance with our SED and FEROS spectrum analysis (see below).

We verified that reasonable changes in the adopted mass ratio and limb-darkening parameters have negligible impact on the results, leading to much smaller changes in the final parameters than the parameter errors. At the same time, the small influence of these parameter changes on the $\chi^2$ value prevents their precise determination from the fit.


\section{Results}\label{sec:results}

In this section, we analyze the multi-frequency SED of HD 352 along with its high-resolution FEROS spectrum to derive the stellar parameters of the red giant. These parameters are then incorporated into the interferometric model to determine the angular diameter of the star, orientation on the sky across various orbital phases, orbital inclination, and linear radius. We then use these results to estimate the masses and mass ratio of the components and to discuss the evolutionary stage of the system.

\subsection{Stellar parameters}

Creating a realistic model requires having proper estimates of the coefficients to be used for the limb darkening. This, in turn, requires knowing the effective temperature and gravity, $\log g$, of the star. 

We first analyzed the multi-frequency SED of HD~352. The parameters of the best-fitting empirical spectrum are (see Fig. \ref{fig:sed}): T$_{\rm eff}$ = 4\,000\,$\pm$\,50 K (with T$_{\rm eff}$ = 4\,026 K from the polynomial fit to the T$_{\rm eff}$--$\chi^2$ dependence), $\log g$ = 1.0\,$\pm$\,0.329 dex (with $\log g$ = 1.15 dex from the $\log g$--$\chi^2$ dependence), and metallicity [M/H] = $-1.0\,\pm\,0.25$ dex. The best-fitting value for the extinction is A$_{\rm V}$ = 0.03~mag.

We note that in general $\log g$ and [M/H] are not well constrained from the SED; for instance, the low metallicity is particularly driven by the bluest datapoints, and while the best-fit temperature remains unchanged if they are omitted, the $\log g$ and [M/H] values vary. Any change in temperature leads to a significantly worse fit, which is not true for the other parameters. The photometric observations used in this analysis were not obtained simultaneously (and not necessarily at the same orbital phase), but the amplitude of the photometric variability is too small to significantly influence the results. 

The comparison of the SED with the best-fitting theoretical spectrum also indicates that the companion does not contribute significantly to the studied wavelength range. Unfortunately, flux measurements in the ultra-violet (UV), where the contribution of the giant is minimal, are not available.

We then used the effective temperature from the SED to obtain more precise stellar parameters from the archival FEROS high-resolution spectrum of HD~352. The best-fitting synthetic spectrum (see Fig. \ref{fig:spectrum}) yielded the following stellar parameters: T$_{\rm eff}$ = 4\,000 K (fixed), $\log g$ = 1.27\,$\pm$\,0.06 dex, and [M/H] = -0.13\,$\pm$\,0.03 dex. Although the formal errors are small, the best values slightly differ depending on the continuum normalization, spectral regions used for the analysis, code, model atmospheres, or solar abundances used for the synthesis. The $v\sin i$ of the spectrum was fixed to the value obtained from the FEROS cross-correlation function calibration to $v\sin i$ by  \citet{2001A&A...375..851M}. We obtained $v\sin i$ = 27.29\,$\pm$\,0.65 km/s, slightly higher (but fully compatible and more precise) than the value of $v\sin i$ = 22\,$\pm$\,3 km/s reported by \citet{1986IBVS.2952....1E}.

\begin{figure}[]
\centering
\includegraphics[width=\columnwidth]{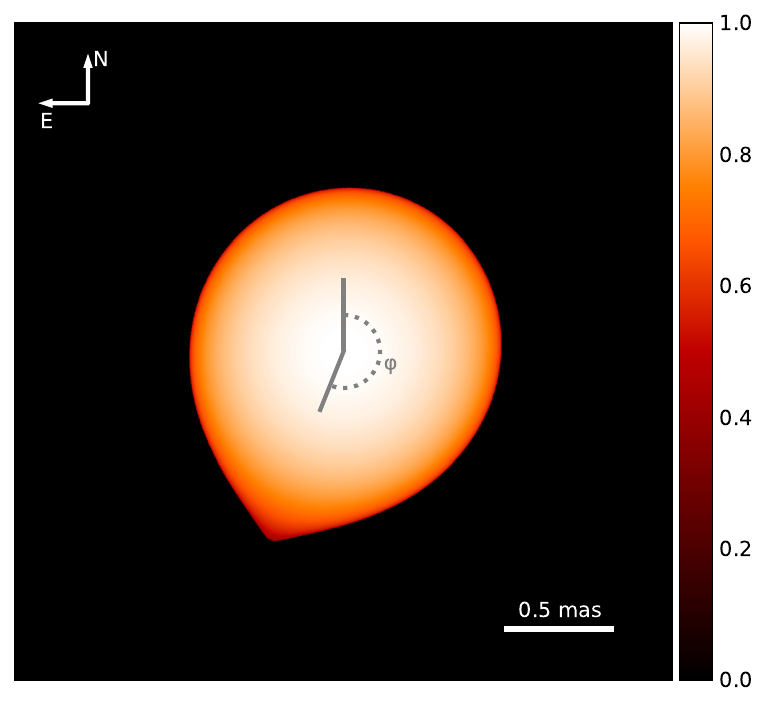}
\caption{The model image used for fitting the interferometric observations. This example presents the best-fitting model for the VLTI/PIONIER observation obtained on 12 December 2020. The two fitted free parameters were the scaling factor of the model image (tied to the angular size of the star) and the orientation of the object ($\varphi$).}\label{fig:model}
\end{figure}

\begin{table*}[h]
\caption{Results of the analysis of interferometric observations of HD~352.}\label{tab:results}
\begin{tabular*}{\textwidth}{@{\extracolsep\fill}lccccccccc}
\hline
\noalign{\smallskip}
& \multicolumn{4}{@{}c@{}}{Limb-darkened disk} & \multicolumn{5}{@{}c@{}}{Roche-lobe filling star} \\%
Night & D [mas]  & V$^2$ $\chi^2$ & CP $\chi^2$ & $\chi^2$ & D [mas] & $\varphi$ [\textdegree] & V$^2$ $\chi^2$ & CP $\chi^2$ & $\chi^2$ \\
\hline
\noalign{\smallskip}
04 December 2010  & 1.49\,$\pm$\,0.01 & 0.69 & 1.50 & 0.89 & 1.47\,$\pm$\,0.02 & 188.7* & 0.64 & 1.56 & 0.87\\
13 August 2012  & 1.56\,$\pm$\,0.01 & 1.12 & 2.95 & 1.58 & 1.53\,$\pm$\,0.01 & 43.9\,$\pm$\,2.2 & 0.49 & 1.67 & 0.79\\
15 October 2013  & 1.50\,$\pm$\,0.01 & 0.83 & 1.95 & 1.11 & 1.49\,$\pm$\,0.01 & 243.5* & 0.68 & 1.56 & 0.90\\
22 September 2014  & 1.44\,$\pm$\,0.01 & 2.83 & 8.46 & 4.24 & 1.44\,$\pm$\,0.01 & 47.2\,$\pm$\,1.0 & 1.90 & 7.80 & 3.37\\
22 December 2019  & 1.37\,$\pm$\,0.02 & 0.43 & 2.78 & 1.02 & 1.36\,$\pm$\,0.02 & 91.2* & 0.43 & 2.77 & 1.02\\
12 December 2020  & 1.43\,$\pm$\,0.01 & 1.75 & 6.86 & 3.03 & 1.38\,$\pm$\,0.01 & 201.4\,$\pm$\,1.7 & 0.83 & 1.95 & 1.11\\
18/19 December 2020  & 1.29\,$\pm$\,0.01 & 1.46 & 0.95 & 1.33 & 1.26\,$\pm$\,0.01 & 178.0* & 1.28 & 1.01 & 1.22\\
\hline
\end{tabular*}

\tablefoot{*The orientation angle was fixed in the fitting procedure. To obtain the final values, weights of 0.5 for the closure phases (CP) and 1.0 for squared visibilities (V$^2$) were adopted (see text). The total $\chi^2$ given in the table corresponds to such a fitting procedure. The uncertainties of diameter (D) and orientation angle ($\varphi$) were obtained using the bootstrapping method.}
\end{table*}

\begin{figure}[]
\centering
\includegraphics[width=\columnwidth]{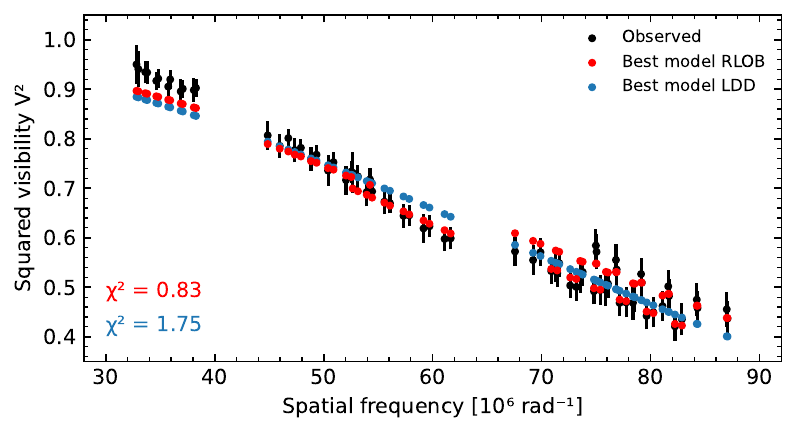}
\caption{VLTI/PIONIER squared visibilities observed on 12 December 2020 as a function of spatial frequency. The observed data are displayed in black with corresponding error bars. The best-fitting model of the deformed star is shown in red ($\chi^2$ = 0.83), and the best-fitting limb-darkened disk model is shown in blue ($\chi^2$ = 1.75). The same limb-darkening law is applied in both cases.}\label{fig:observables_v2}
\end{figure}

\begin{figure}[]
\centering
\includegraphics[width=\columnwidth]{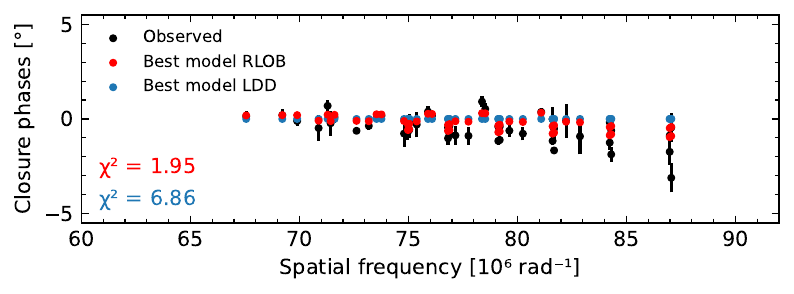}
\caption{VLTI/PIONIER closure phases observed on 12 December 2020 as a function of spatial frequency (the longest baseline in the triangle). The colors are same as in Fig. \ref{fig:observables_t3phi}.}\label{fig:observables_t3phi}
\end{figure}

\subsection{Angular diameter and orientation from interferometry}
Having inferred the stellar parameters, we created a model image of the Roche-lobe filling star with limb-darkening (Fig.~\ref{fig:model}; Sec. \ref{sec:methods_model}). This model image was then used to fit the interferometric observations. Two free parameters were fitted using our Julia script employing OITOOLS for the calculation of interferometric observables: the scaling factor of the image (tied to the Roche-lobe radius) and the orientation of the star (angle $\varphi$ clockwise from north to the direction towards the L1 point; see Fig.~\ref{fig:model}).

During the $\chi^2$ minimization procedure, we initially considered only the squared visibilities, but we also tested simultaneous minimization including closure phases (with weights of either 0.5 or 1.0). For most observations, this did not significantly alter the values of the two fitted parameters, except in cases with only shorter baselines available (e.g., observations from 2010). For the final analysis, we assigned a weight of 0.5 to the closure phases. Using a higher weight led to a minor improvement in the total $\chi^2$ only (due to the relatively large uncertainties in the closure phases), but significantly increased the $\chi^2$ for the squared visibilities and led to inconsistencies in the derived parameters. For example, the orientation of the star became constrained primarily by overfitting the uncertain closure phases rather than being constrained by the squared visibilities, resulting in orientations far from those expected from the orbital phase.

In some cases, the orientation of the star was poorly constrained from fitting a single observation. To address this, we determined the orientations we determined the orientation for the observations on 13 August 2012 and 12 December 2020 simultaneously, enforcing the difference between the two angles to the expected value based on the difference in orbital phases. We used the orbital parameters of \citet{2011MNRAS.410.1761K} -- orbital period of 96.4371 days, T$_0$ (HJD) = 2\,453\,602.1473 -- to calculate the orbital phase. We then used these values to calculate the expected orientation for the remaining epochs. Generally, for observations with longer baselines, the expected values calculated in this manner did not significantly differ from the fitted values when the orientation was kept as a free parameter.

The fitted scaling factor of the model image and the adopted distance of 361.3\,$\pm$\,8.8 pc, obtained from \textit{Gaia} DR3 parallax \citep{2023A&A...674A...1G}, allowed us to calculate the radius of the star for each epoch. The typical error of the radius for an individual epoch is about 1.5 - 2~R$_\odot$. 
The average value from seven epochs of interferometric observations is 54.5\,$\pm$\,2.7~R$_\odot$ (orientation angles fixed by 2012 and 2020 observations and a weight of 0.5 for the closure phases in the fitting procedure).

\subsection{Inclination and the projection effect on interferometry}

Most of the available interferometric observations were obtained very close to quadrature. This means the projection would have only a minimal impact on the results. The impact would be largest near conjunctions, and its amplitude depends on the inclination of the system (no impact for $i$ = 0\textdegree, but the star would appear effectively circular at conjunction for $i$ = 90\textdegree).

Given that the inclination of the system is a priori unknown, we estimated it from the rotational velocity $v\sin i$, making the reasonable assumption for such a short-period binary with a red giant that the rotation of the giant is synchronized with the orbital period. This estimate requires, in addition to $v\sin i$ that we have from the FEROS spectrum, knowledge of the radius of the star. We used the average value obtained from the interferometric measurements in the previous step. This led to an inclination of 72.6\textdegree$^{+17.4}_{-10.1}$.

We adopted this inclination and, for each epoch of interferometric observation, calculated the projection of our Roche-lobe-filling star. We then used a model image tuned for each observation and fitted the interferometric observables as described in the previous section. The obtained value of the radius was used to reevaluate the inclination, and the procedure was repeated again.

In general, the inclusion of the projection effect had minimal impact on the shape of the star, its diameter, and orientation for most epochs, except for the observations from 15 October 2013 and 22 December 2019. In particular, during the latter observation, the system was near conjunction (orbital phase 0.06), which made the star appear nearly circular. The total $\chi^2$ values for the fits with and without projection were comparable, largely due to the relatively high uncertainties in the interferometric observables (the squared visibilities in both cases and closure phases for the 2013 epoch; see Figs. \ref{fig:interf_2013}, \ref{fig:interf_2019_2}, \ref{fig:interf_2013_CP}, and \ref{fig:interf_2019_2_CP}) and the fact that the deformation is affecting more significantly the longest baselines. This is especially the case of the 2019 observations, for which the diameter of the star differed by approximately 5\% -- the star appears smaller when the projection effect is not included.

The resulting radius of the giant in HD~352 was 55.0\,$\pm$\,2.6 R$_\odot$ and the corresponding inclination was 71.0\textdegree$^{+19.0}_{-9.2}$. The results for the individual epochs are summarized in Table~\ref{tab:results}. An example of the fit to the squared visibilities and closure phases (for the observations obtained on 12 December 2020; orbital phase 0.76) is shown in Fig. \ref{fig:observables_v2} and \ref{fig:observables_t3phi}. 

\subsection{Comparison with limb-darkened disk model}

For comparison, we used the same code as for our Roche-lobe model to fit the interferometric observations with a symmetric disk model, adopting the same limb-darkening law. In this case, the single fitted parameter was the scaling factor, as the angle is meaningless for the symmetric disk. In all but one case, the $\chi^2$ values were worse compared to our Roche-lobe-filling giant model, with significant differences, especially for those observations obtained with the longest baselines near the quadrature of the system. The best example is the dataset obtained on 12 December 2020 (see Table \ref{tab:results}), where the $\chi^2$ of the symmetric disk is almost three times the one associated with the Roche-lobe model. As a check of our code, we fitted the limb-darkened disk model to observations also using the PMOIRED code \citep{2022SPIE12183E..1NM} and obtained identical diameters.

\subsection{Mass ratio and individual masses}

Using the inclination derived from interferometric observations, the spectroscopic mass function of \citet{2011MNRAS.410.1761K}, and assuming that the giant fully fills its Roche lobe (thus, the Roche lobe radius calculated using  the formula of \cite{1983ApJ...268..368E} is equal to the radius of the star, 55.0\,$\pm$\,2.6 R$_\odot$, inferred from interferometry), we calculated the masses of the components of the system and their mass ratio. The obtained masses are as follows: M$_{\rm G}$ = 1.97$^{+0.30}_{-0.27}$ M$_\odot$, M$_{\rm h}$ = 1.16$^{+0.22}_{-0.16}$ M$_\odot$, and q = M$_{\rm G}$/M$_{\rm h}$ = 1.69$^{+0.22}_{-0.23}$.  

If we calculate the mass of the star from the interferometric radius and $\log g$ from spectroscopy (in principle inclination independent), we obtain M$_{\rm G}$ = 2.06$^{+0.53}_{-0.43}$ M$_\odot$, with the uncertainties being primarily influenced by the uncertainty in the radius.

\begin{figure}[]
\centering
\includegraphics[width=\columnwidth]{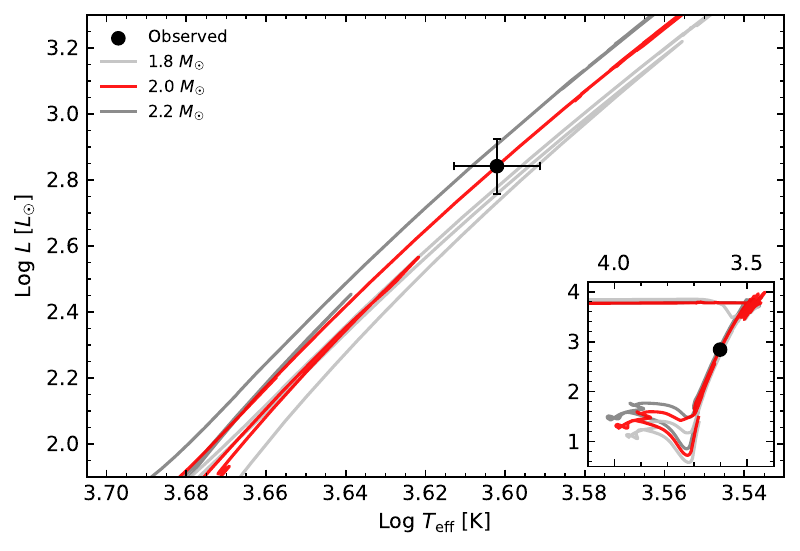}
\caption{Position of the cool component of HD~352 in the HR diagram (black symbol). The MESA evolutionary tracks \citep{2011ApJS..192....3P} obtained using the MIST Web Interpolator \citep{2016ApJS..222....8D,2016ApJ...823..102C} for initial masses of 1.8, 2.0, and 2.2 M$_\odot$ are shown in light gray, red, and dark gray, respectively. A metallicity of -0.13 dex was adopted. The inset shows the whole isochrones. }\label{fig:hr_diag}
\end{figure} 

\subsection{Position in the Hertzsprung–Russell diagram}

The inferred temperature and radius allowed us to calculate the luminosity of the giant in HD~352 using the Stefan-Boltzmann law. The temperature and luminosity were then used to place HD~352 into the HR diagram (see Fig. \ref{fig:hr_diag}). Its position was compared with the stellar evolutionary tracks from the Modules for Experiments in Stellar Astrophysics \citep[MESA;][]{2011ApJS..192....3P}, obtained using the MIST Web Interpolator \citep{2016ApJS..222....8D,2016ApJ...823..102C}, confirming that the derived values are compatible with 2.0 $\pm$ 0.2 M$_\odot$ tracks. A metallicity of -0.13 dex, derived from the analysis of the FEROS spectrum, was adopted.  

\section{Discussion}\label{sec:discussion}
\subsection{Detection of the tidal deformation via interferometry}
HD~352 was observed on 10 occasions with VLTI/PIONIER between 2010 and 2020. Several of the observations were obtained with sufficiently long baselines, allowing the shape of the donor in this binary system to be resolved. We have demonstrated that observations near quadrature cannot be explained by simple symmetric disk models but are consistent with the shape of a Roche-lobe-filling star (see Table \ref{tab:results}). To our knowledge, this is the first case of the tidal deformation of a red giant star being observed directly using interferometric techniques. The only comparable result is the observation of $\beta$~Lyrae \citep{2008ApJ...684L..95Z,2018A&A...618A.112M,2021A&A...645A..51B}. However, this system is very different from HD 352, as both components are hot B-type stars, while the donor is less massive than the accretor, and the environment -- and consequently the interferometric observables -- are more complex.

The interferometric detection of the tidally deformed star in the HD~352 system is consistent with previous observations, particularly the photometric variability recurring with half of the spectroscopically confirmed orbital period, explained as ellipsoidal variability. Additionally, as stated in the introduction, the deformation of the giant is likely indicated by the small but non-zero apparent eccentricity of the orbital solution based on very precise radial velocity monitoring. These observations support the model we created to fit the interferometric observables, which assumes that the giant is fully filling its Roche lobe.

From interferometry, we can infer the shape and the size of the giant star. Thus, using our model that incorporated the projection effect and eight interferometric datasets\footnote{With two of them being analyzed together and an additional two not being used due to their short baselines.}, we derive a radius of the giant of 
 55.0\,$\pm$\,2.6~R$_\odot$. Such a radius further supports the assumption of the giant filling its Roche lobe. The lower limit for the star radius is given by the fact that $v\sin i$ should always be smaller or equal to the rotation velocity, v$_{\rm rot} = 2 \pi R/P_{\rm orb}$, if the rotation is synchronized with the orbital period, which is a reasonable assumption for HD~352. For that reason, the radius of the giant cannot be smaller than 52~R$_\odot$ for the $v\sin i$ obtained from the FEROS archival spectrum. This strongly suggests that it is the giant itself in the interferometric observations, not its wind, that is filling the Roche lobe. Assuming that stellar rotation axis is perpendicular to the orbital plane, and using our estimated radius value, we indeed recover exactly our measured $v \sin i$ value. 

The exact value of the inferred radius of the star depends on several factors, including the adopted limb-darkening law, mass ratio adopted for the model, and other parameters such as the adopted distance (obtained from \textit{Gaia} parallax in our case; the parallax over error is $\sim$41) or the inclination used in the modeling. However, we tested that reasonable changes in the fixed parameters do not significantly influence any results and have no impact on our conclusions. It should also be noted that in some cases, our model does not provide an ideal fit to the interferometric observables, particularly the squared visibilities at the shortest available baselines. One possible explanation is the presence of a very faint but inhomogeneous background. Including some background flux with absorption regions improved the fit at the shortest baselines in our LITpro analysis; however, the data are not sufficient to constrain the parameters of these structures from the fit, so we did not include such background in the final analysis.

\subsection{Masses, mass ratio, and the evolutionary perspective}

With the precise spectroscopic mass function from \cite{2011MNRAS.410.1761K}, we were able to use the interferometric radius of the star, combined with the inclination of the system, to obtain the masses of the individual components and their mass ratio. The giant's mass obtained in this way, M$_{\rm G}$ = 1.97$^{+0.30}_{-0.27}$ M$_\odot$, is within the errors consistent with the mass obtained from the interferometric radius and $\log g$ from spectroscopy (M$_{\rm G}$ = 2.06$^{+0.53}_{-0.43}$ M$_\odot$), which is, in principle, independent of the inclination. 
Additionally, the position of the star in the HR diagram is consistent with the evolutionary track of a star with an initial mass of 2.0 M$_\odot$ and with the metallicity of HD~352 inferred from spectroscopy (Fig. \ref{fig:hr_diag}). The mass from the evolutionary track, therefore, agrees with the other two estimates derived in this work, within their respective uncertainties. The location in the HR diagram, when compared with the MESA evolutionary tracks, also suggests that the star is already on the asymptotic giant branch (AGB) but has not yet experienced thermal pulses. This AGB classification is further supported by the fact that the star is filling its Roche lobe. According to the MESA calculations, a 2.0 M$_\odot$ star would fill only about two-thirds of the current Roche lobe at the tip of the first red giant branch.

The mass of the companion is M$_{\rm h}$ = 1.16$^{+0.22}_{-0.16}$ M$_\odot$. Such a mass is consistent with a F8V star\footnote{Spectral types between G2V and F4V can be accommodated within the errors.)} \citep{2013ApJS..208....9P}, and this result is consistent with the UV study of \citet{1988AcA....38..353E}, who claimed that the low-resolution IUE spectra suggest the presence of a mid-F accretor.

The interferometric radius of the giant, assuming the giant is filling the Roche lobe, also constrains the mass ratio. We obtained q = M$_{\rm G}$/M$_{\rm h}$ = 1.69$^{+0.22}_{-0.23}$. An independent constraint on the mass ratio can be provided by the photometric variability. The mass ratio is bound to inclination through the spectroscopic mass function. Given that the inclination influences the observability of ellipsoidal variability, the observed amplitude of about 0.17 - 0.20 mag in the photometric observation from the Multi-site All-Sky CAmeRA (MASCARA) survey \citep[unfiltered data with peak quantum efficiency of the camera at $\sim$ 5\,000\,\AA;][]{2018A&A...617A..32B} and from the Hipparcos epoch photometry \citep[$\lambda_{\rm eff}$ = 4\,902\,\AA;][]{1997ESASP1200.....E}, provide a constraint on these two parameters. Our calculations suggest that for inclinations $\lesssim$ 62\textdegree\,\,(i.e., mass ratio q $\lesssim$ 1.54), the expected amplitude would be much smaller than the observed one for the adopted stellar parameters and limb-darkening law. For example, to reach q = 1.3, the inclination needs to be $\sim$ 50\textdegree, but the amplitude of the variability would be $<$ 0.10 mag.

The  value we obtain for the mass ratio ($q$ = 1.69) is well above the limit considered necessary for stable mass transfer \citep[see, e.g., the discussion in][]{2023A&A...669A..45T}. However, recent studies on the stability of mass transfer in binaries \citep[e.g., ][]{2002ApJ...565.1107P, 2010ApJ...717..724G, 2015ApJ...812...40G, 2020ApJ...899..132G} suggest that the range of mass ratios and orbital periods allowing stable mass transfer might be larger than previously predicted by classical studies. In a very recent work, \citet{2023A&A...669A..45T} proposed additional physically motivated instability criteria and found increased stability of mass transfer in convective giants, claiming that the critical mass ratio increases from about 1 at the base of the RGB to about 10 at the onset of thermal pulses on the AGB. 
Our inferred mass ratio is still above the limiting value for a self-regulated stable mass transfer, but taking into account the fact that the metallicity of HD 352 is slightly sub-solar while the models are for $\sim$ solar metallicity, within the errors of the masses and interferometric radius, it might not lie far. 

If the mass ratio is indeed greater than the critical mass ratio, this suggests that HD 352 is now in a very short-lasting evolutionary stage during which the Roche lobe shrinks in response to mass loss, while the giant's radius increases, leading to unstable mass transfer. Consequently, the system will soon enter the common envelope phase (\citeauthor{1976IAUS...73...75P} \citeyear{1976IAUS...73...75P}; see also \citeauthor{2020cee..book.....I} \citeyear{2020cee..book.....I}), and for a brief period of time, it will be observable as a planetary nebula with a binary post-common envelope central star \citep{2017NatAs...1E.117J}. It should be noted that the probability of finding a system in this very brief stage of evolution is small, given its short duration \citep[up to a few tens or hundreds of thousands of years from the moment when the mass transfer rate exceeds the rate at which wind mass loss occurs to the common envelope evolution;][]{2023A&A...669A..45T}.

However, it is not possible to rule out that HD 352 is undergoing stable mass transfer at this point, even if the models fail to predict this scenario. The incompleteness of our understanding of mass transfer is well documented by several groups of interacting and post-interacting (post-AGB) systems, such as symbiotic binaries or barium stars, which consist of systems containing a red giant and a compact accretor \citep[white dwarfs or neutron stars;][]{2012BaltA..21....5M}. Models predict populations with either long orbital periods (greater than a few thousand days), possibly with larger eccentricities due to orbit widening from wind mass transfer, or populations with very short periods (less than a few days) on circular orbits resulting from the common envelope evolution, with a clear gap expected between these two populations dominated by each formation channel \citep[see, e.g.,][]{2010A&A...523A..10I,2012MNRAS.423.2764N}. However, the observed eccentricity-period distributions of the aforementioned binary classes deviate significantly from these predictions, as many systems with orbital periods in the range of hundreds to a few thousand days exist precisely in the period gap \citep[e.g.,][]{2012BaltA..21....5M, 2019A&A...626A.127J,2020A&A...639A..24E}. Additionally, many objects with periods shorter than 4000 days, while expected to have circular orbits, exhibit mild to high eccentricities \citep[see also][]{2024A&A...682A...7B}. Consequently, both possible scenarios for HD 352, whether it is undergoing stable or unstable mass transfer, have important implications for modeling a large class of binary systems. If the system is undergoing dynamically unstable mass transfer, we are seeing HD~352 in a very transient phase, and further monitoring of the system -- photometric and spectroscopic -- is warranted, as we can expect the star to rapidly show changes in its luminosity and the system to see its orbital period decrease.


\subsection{Implications for interferometric observations of other binaries}

The analysis of the interferometric observations presented in this work has significant implications for other types of interacting binaries, such as symbiotic stars. Beyond the aforementioned discrepancies in orbital periods and eccentricities, the mechanism responsible for the mass transfer from the giant to its companion remains unclear. It is uncertain whether this process is dominated by Roche-lobe overflow, stellar wind, or a combination of both \citep[e.g., wind Roche-lobe overflow; ][]{2007ASPC..372..397M}, although it appears that many symbiotic systems possess accretion disks around the accretors \citep{2024A&A...683A..84M}.

As we have demonstrated, under favorable conditions (observations obtained near quadrature for a star with a sufficiently large angular diameter), it is possible to detect the tidal deformation of the star if it is filling its Roche lobe. Therefore, observing a substantial sample of close symbiotic binaries with interferometry can provide insights into the mass-transfer mechanisms dominating these systems. To date, such observations have been presented for only four symbiotic stars: AG Peg, FG Ser, ER Del, and V1261 Ori \citep{2014A&A...564A...1B}. We have collected and analyzed VLTI/PIONIER and CHARA/MIRC-X observations of a significantly larger sample of symbiotic stars and will present them soon (Merc et al., in prep.).

\section{Conclusions}
In this study, we analyzed VLTI/PIONIER interferometric observations of HD 352, a binary system consisting of a red giant and a main sequence companion in a 96-day orbit. These observations were collected over 10 individual nights across a 10-year period, capturing various orbital phases of the system.

Simple symmetric disk models could not account for the observations, especially those near quadrature phases. To address this, we created a model image of a Roche-lobe-filling star and simulated the resulting interferometric observables. This model successfully matched the interferometric data. The fact that the giant in HD352 is filling its Roche-lobe is also supported by the observed photometric variability, and the spurious slight non-zero eccentricity of the orbit. We believe this represents the first direct confirmation of tidal deformation in a red giant through interferometric techniques.

By combining our interferometric analysis results with stellar parameters derived from high-resolution spectroscopy, SED modeling, and the \textit{Gaia} distance, we determined the radius of the red giant ($\sim$55 R$_\odot$), luminosity ($\sim$700 L$_\odot$), and estimated mass ($\sim$2 M$_\odot$). The precise spectroscopic mass function further allowed us to estimate the mass of the companion ($\sim$1.2 M$_\odot$). The resulting mass ratio suggests that the system may be in a short-lived stage of unstable mass transfer and approaching a common envelope phase. Alternatively, mass transfer might be occurring stably, even though current models do not predict this scenario.

\begin{acknowledgements}
We are thankful to an anonymous referee for the comments and suggestions improving the manuscript. The research of J.M. was supported by the Czech Science Foundation (GACR) project no. 24-10608O. J.M. acknowledges the support received through the on-the-job training programme at the European Southern Observatory (ESO), funded by the Ministry of Education, Youth and Sports of Czechia (MEYS).

Based on observations collected at the European Organisation for Astronomical Research in the Southern Hemisphere under ESO programmes 60.A-9700(A),  086.C-0999(C) (PI: Berger), 089.D-0527(A) (PI: Hillen), 091.D-0344(C) (PI: Jorissen), 093.D-0363(B) (PI: Boffin), 0103.C-0915(C) (PI: Kraus), 0104.C-0737(A) (PI: Kraus), and 0106.C-0880(B) (PI: Kraus).

The Python package PyAstronomy, used in this work, is available at \url{https://github.com/sczesla/PyAstronomy}. The code used for calculating interferometric observables, OITOOLS, can be downloaded from \url{https://github.com/fabienbaron/OITOOLS.jl}.
\end{acknowledgements}

\bibliographystyle{aa}
\bibliography{sn-bibliography.bib}

\begin{appendix}

\section{uv-plane coverage and squared visibilities}
\label{appendix}

\begin{figure}[h]
\centering
\includegraphics[width=0.49\textwidth]{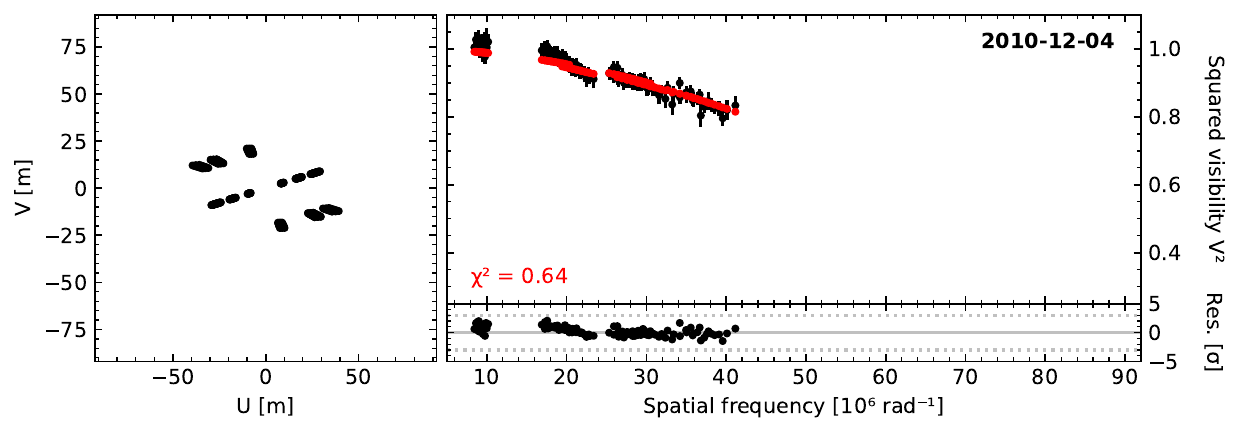}
\caption{VLTI/PIONIER uv-plane coverage (left panel) and squared visibilities as a function of spatial frequency (right panel) observed on 4 December 2010 (orbital phase 0.79). The observed data are shown in black, and the best-fitting model (Roche-lobe filling star; simultaneous fit to squared visibilities and closure phases) is shown in red.}\label{fig:interf_2010}
\end{figure}

\begin{figure}[h]
\centering
\includegraphics[width=0.49\textwidth]{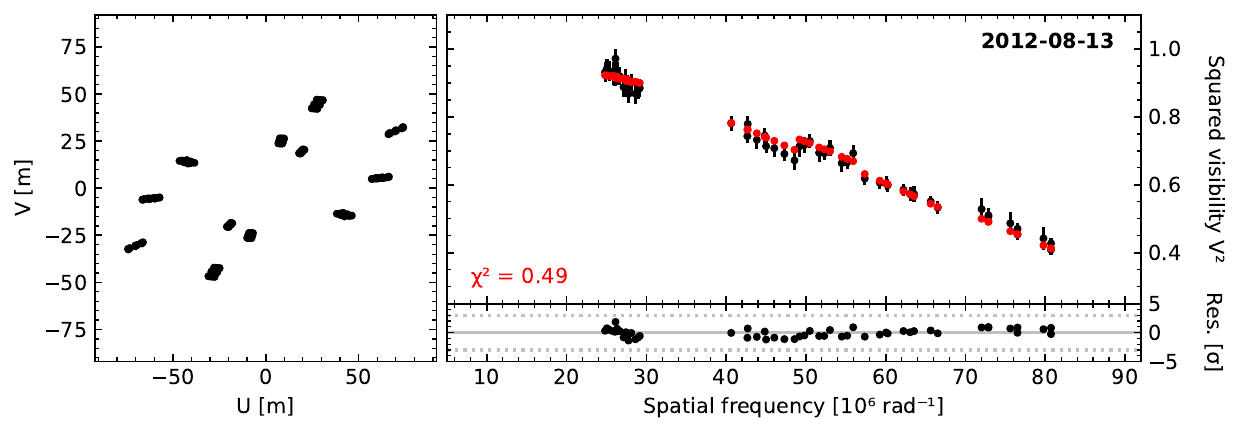}
\caption{Same as Fig. \ref{fig:interf_2010}, but for VLTI/PIONIER observations from 13 August 2012 (orbital phase 0.20).}\label{fig:interf_2012}
\end{figure}

\begin{figure}[h]
\centering
\includegraphics[width=0.49\textwidth]{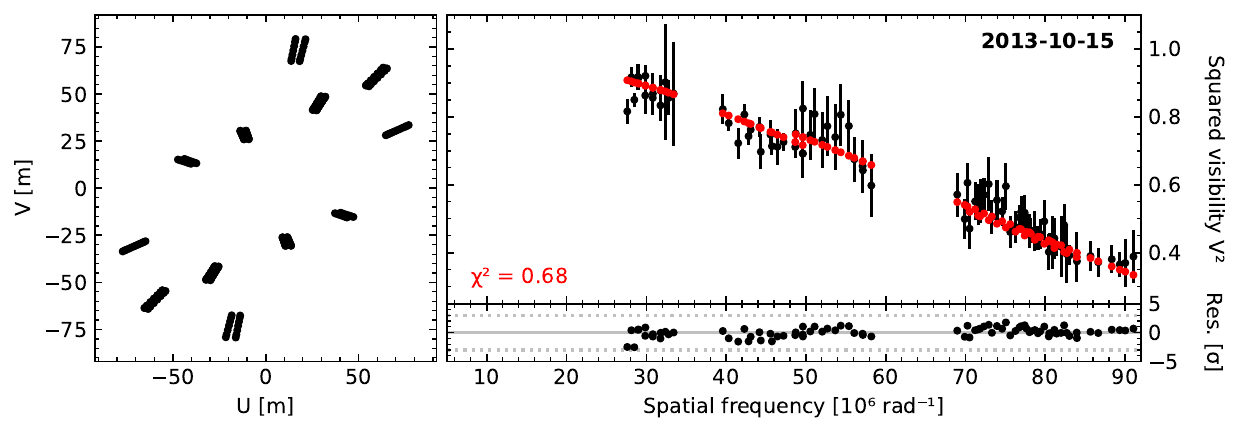}
\caption{Same as Fig. \ref{fig:interf_2010}, but for VLTI/PIONIER observations from 15 October 2013 (orbital phase 0.64).}\label{fig:interf_2013}
\end{figure}

\begin{figure}[h]
\centering
\includegraphics[width=0.49\textwidth]{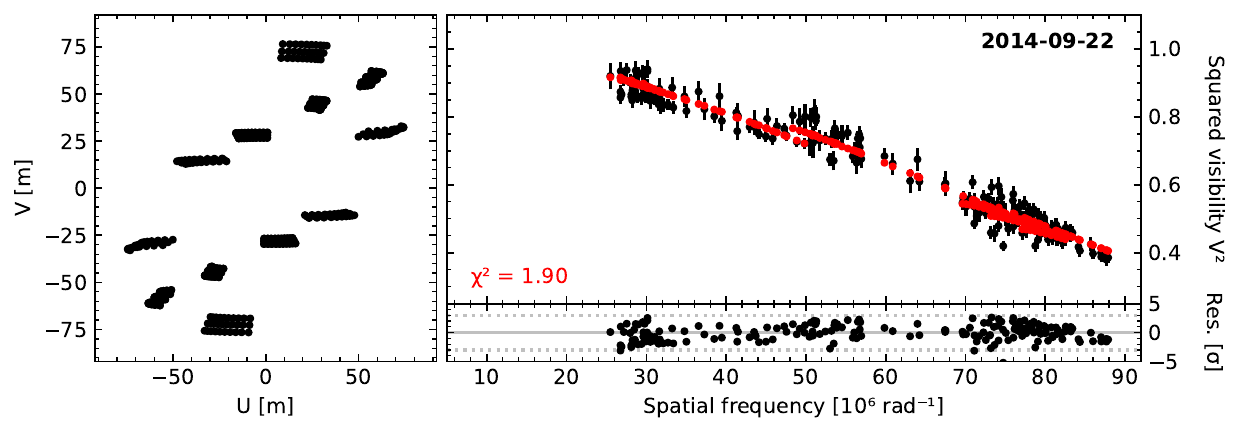}
\caption{Same as Fig. \ref{fig:interf_2010}, but for VLTI/PIONIER observations from 22 September 2014 (orbital phase 0.19).}\label{fig:interf_2014}
\end{figure}

\begin{figure}[h]
\centering
\includegraphics[width=0.49\textwidth]{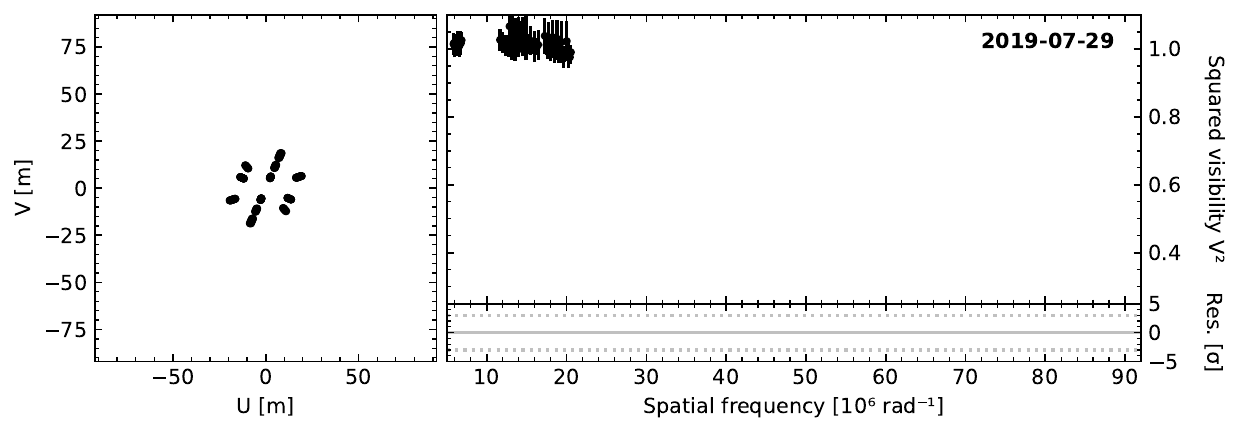}
\caption{Same as Fig. \ref{fig:interf_2010}, but for VLTI/PIONIER observations from 29 July 2019 (orbital phase 0.55).}\label{fig:interf_2019_1}
\end{figure}

\begin{figure}[h]
\centering
\includegraphics[width=0.49\textwidth]{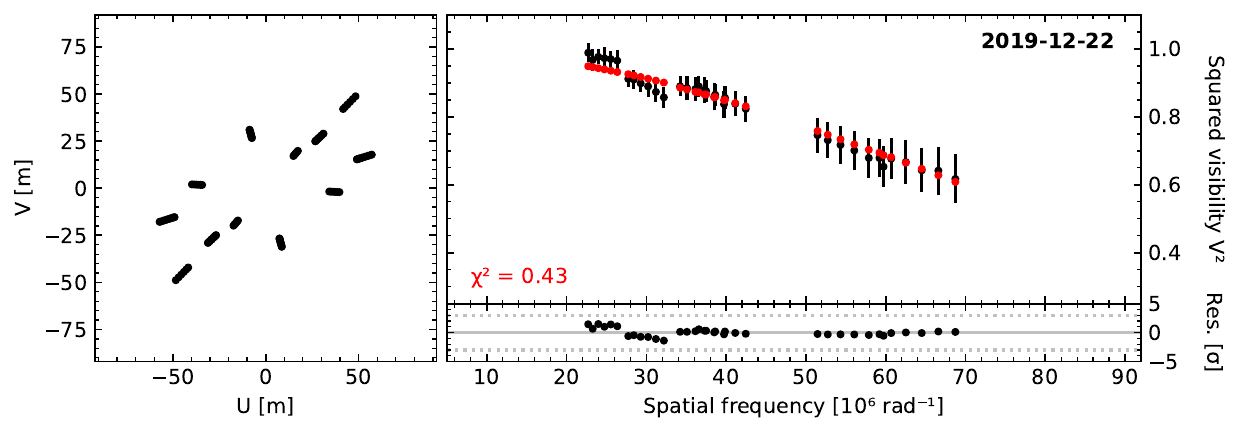}
\caption{Same as Fig. \ref{fig:interf_2010}, but for VLTI/PIONIER observations from 22 December 2019 (orbital phase 0.06).}\label{fig:interf_2019_2}
\end{figure}

\begin{figure}[h]
\centering
\includegraphics[width=0.49\textwidth]{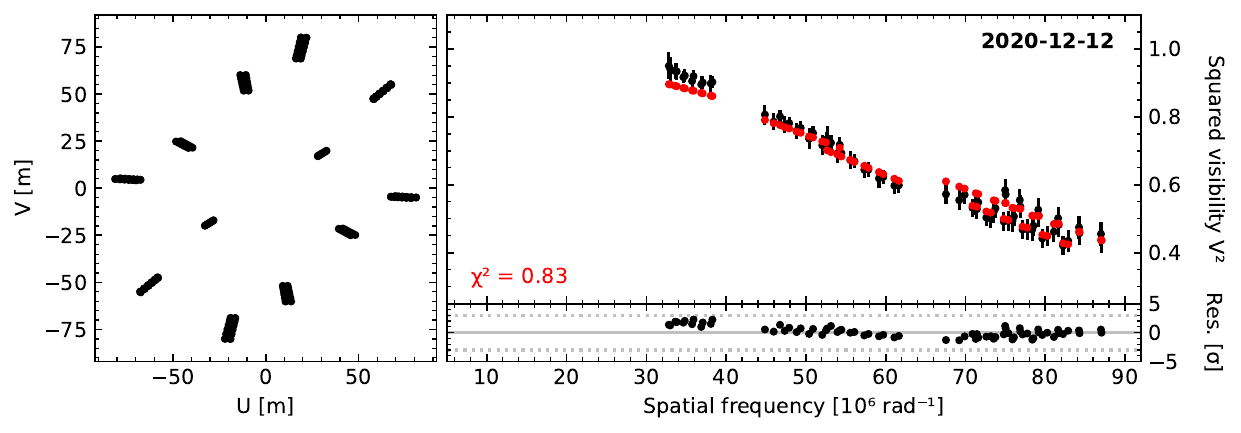}
\caption{Same as Fig. \ref{fig:interf_2010}, but for VLTI/PIONIER observations from 12 December 2020 (orbital phase 0.76).}\label{fig:interf_2020_1}
\end{figure}

\begin{figure}[h]
\centering
\includegraphics[width=0.49\textwidth]{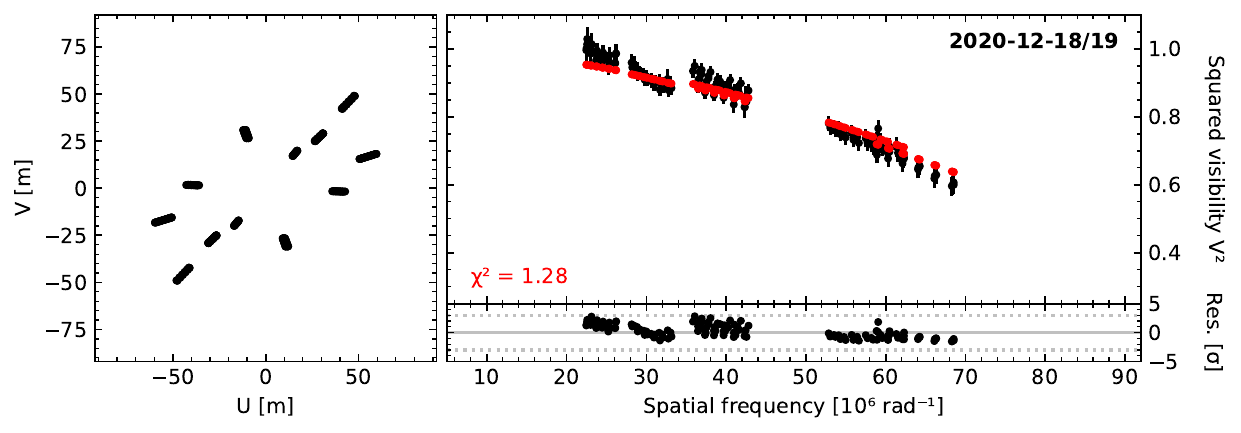}
\caption{Same as Fig. \ref{fig:interf_2010}, but for VLTI/PIONIER observations from 18/19 December 2020 (orbital phase 0.82).}\label{fig:interf_2020_2}
\end{figure}

\begin{figure}[h]
\centering
\includegraphics[width=0.49\textwidth]{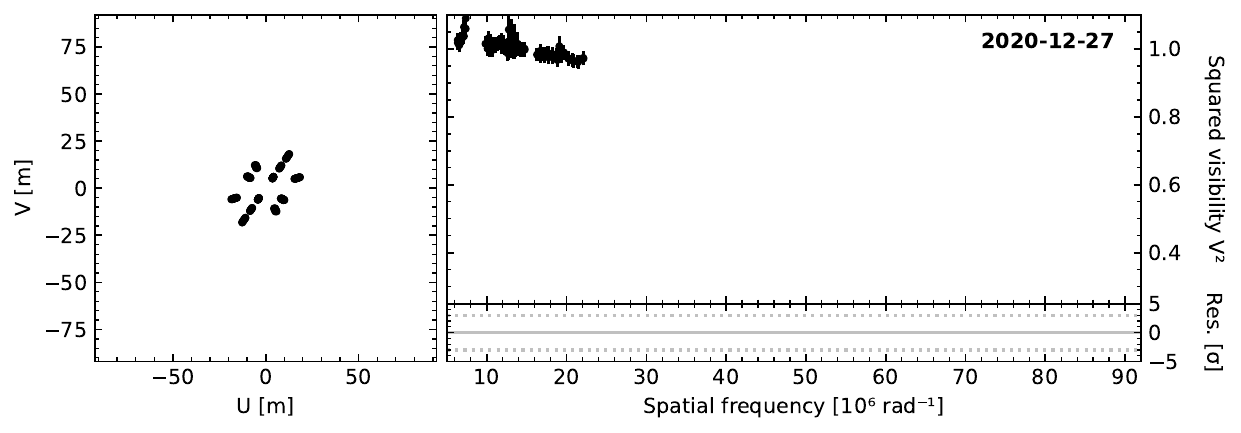}
\caption{Same as Fig. \ref{fig:interf_2010}, but for VLTI/PIONIER observations from 27 December 2020 (orbital phase 0.91).}\label{fig:interf_2020_3}
\end{figure}

\begin{figure}[h]
\section{Closure phases}
\label{appendix_b}
\centering
\includegraphics[width=0.49\textwidth]{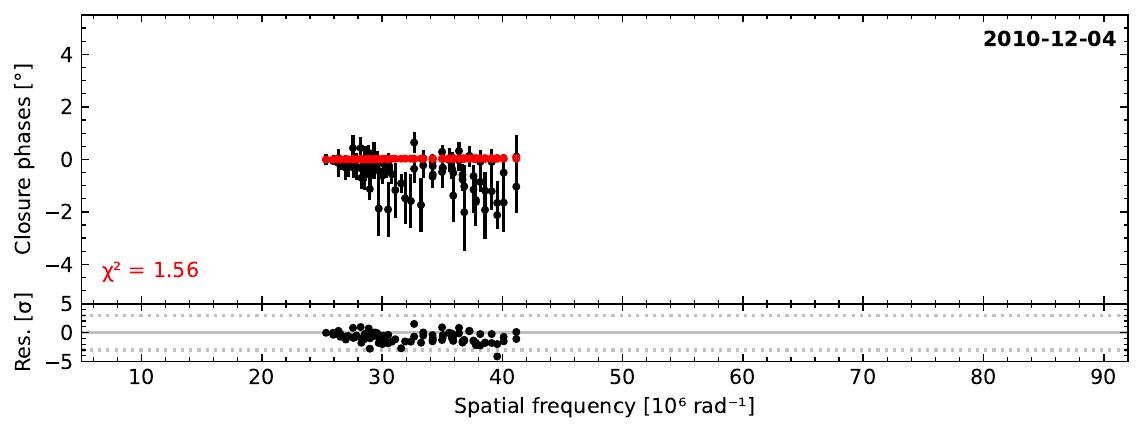}
\caption{Closure phases observed on 4 December 2010 (orbital phase 0.79). The observed data are shown in black, and the best-fitting model (Roche-lobe filling star; simultaneous fit to squared visibilities and closure phases) is shown in red.}\label{fig:interf_2010_CP}
\end{figure}

\begin{figure}[h]
\centering
\includegraphics[width=0.49\textwidth]{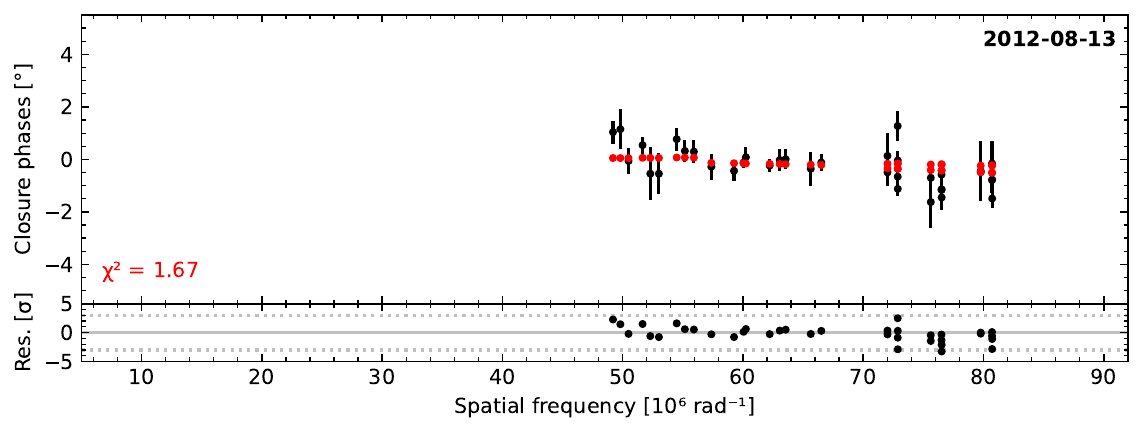}
\caption{Same as Fig. \ref{fig:interf_2010_CP}, but for VLTI/PIONIER observations from 13 August 2012 (orbital phase 0.20).}\label{fig:interf_2012_CP}
\end{figure}

\begin{figure}[h]
\centering
\includegraphics[width=0.49\textwidth]{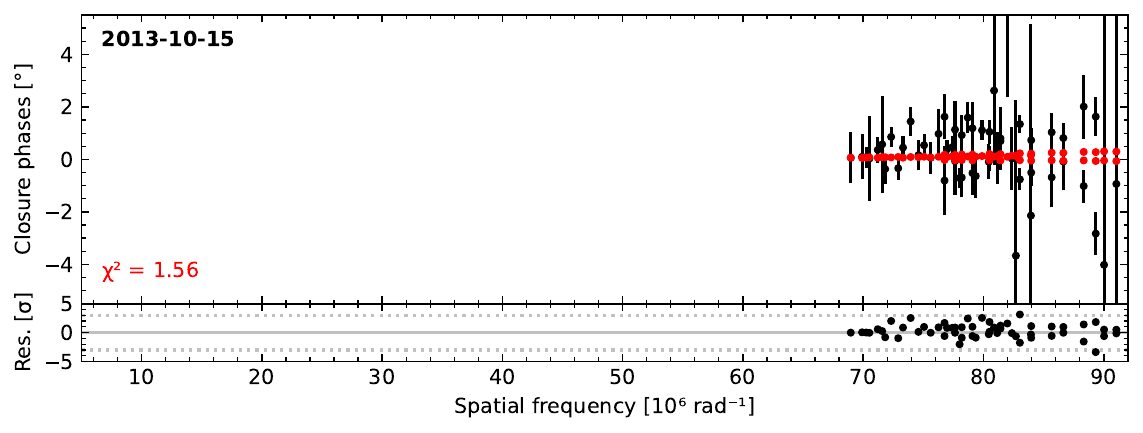}
\caption{Same as Fig. \ref{fig:interf_2010_CP}, but for VLTI/PIONIER observations from 15 October 2013 (orbital phase 0.64).}\label{fig:interf_2013_CP}
\end{figure}

\begin{figure}[h]
\centering
\includegraphics[width=0.49\textwidth]{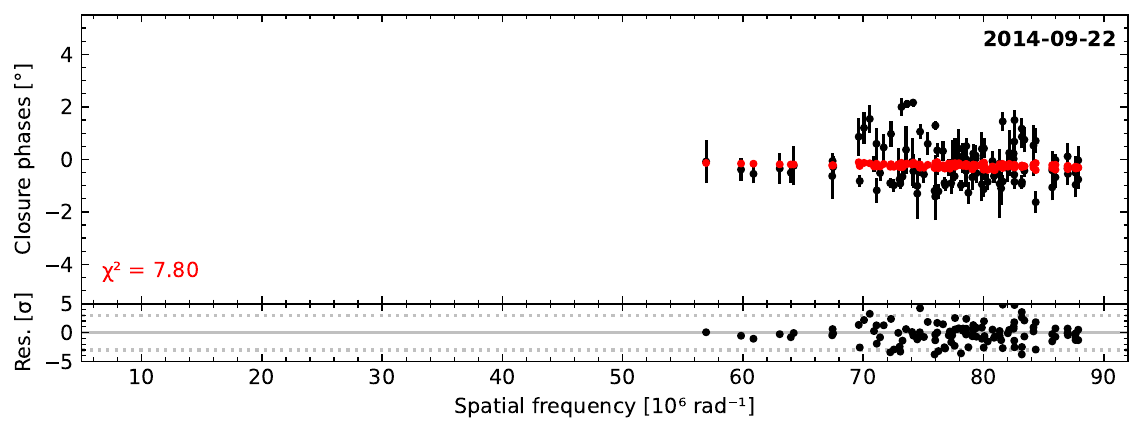}
\caption{Same as Fig. \ref{fig:interf_2010_CP}, but for VLTI/PIONIER observations from 22 September 2014 (orbital phase 0.19).}\label{fig:interf_2014_CP}
\end{figure}

\begin{figure}[h]
\centering
\includegraphics[width=0.49\textwidth]{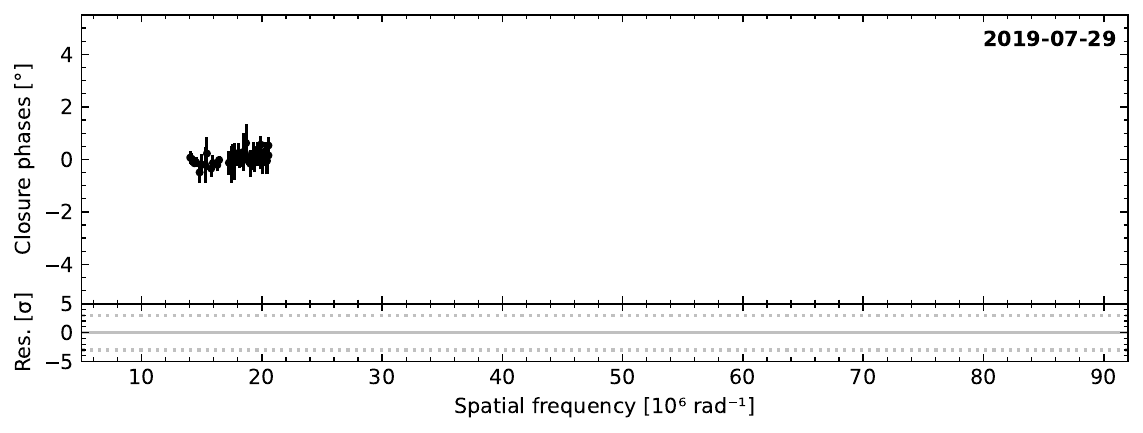}
\caption{Same as Fig. \ref{fig:interf_2010_CP}, but for VLTI/PIONIER observations from 29 July 2019 (orbital phase 0.55).}\label{fig:interf_2019_1_CP}
\end{figure}

\begin{figure}[h]
\centering
\includegraphics[width=0.49\textwidth]{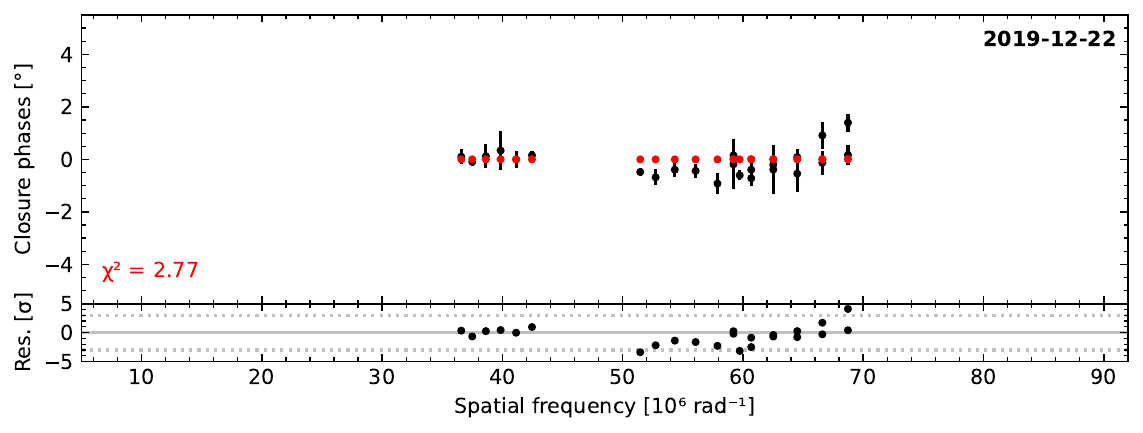}
\caption{Same as Fig. \ref{fig:interf_2010_CP}, but for VLTI/PIONIER observations from 22 December 2019 (orbital phase 0.06).}\label{fig:interf_2019_2_CP}
\end{figure}

\begin{figure}[h]
\centering
\includegraphics[width=0.49\textwidth]{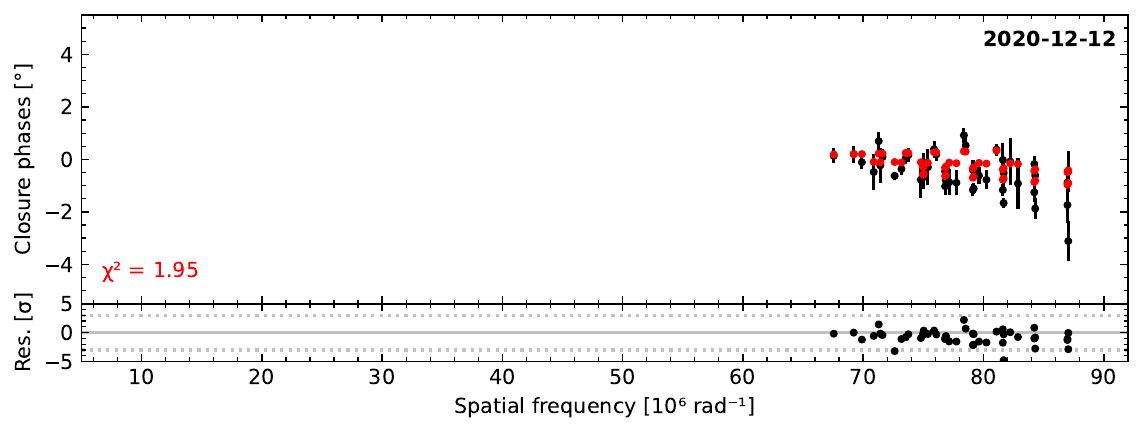}
\caption{Same as Fig. \ref{fig:interf_2010_CP}, but for VLTI/PIONIER observations from 12 December 2020 (orbital phase 0.76).}\label{fig:interf_2020_1_CP}
\end{figure}

\begin{figure}[h]
\centering
\includegraphics[width=0.49\textwidth]{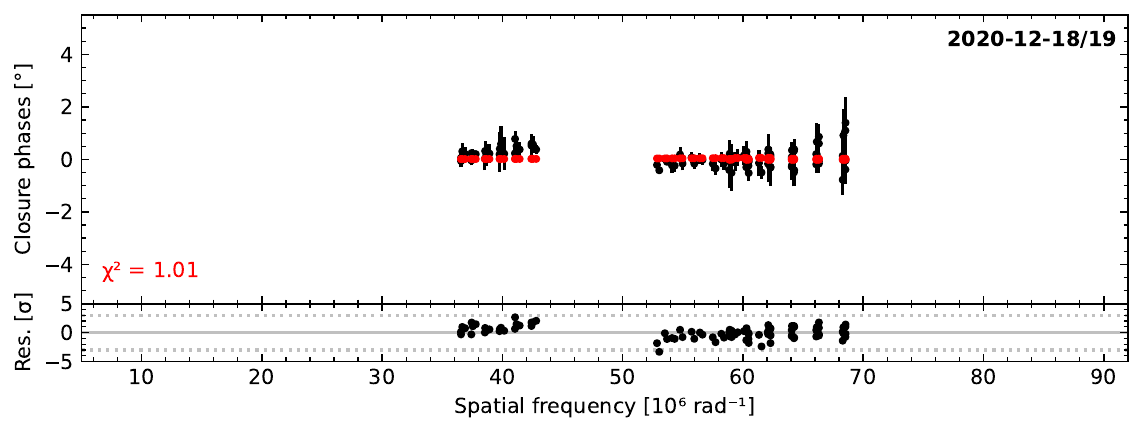}
\caption{Same as Fig. \ref{fig:interf_2010_CP}, but for VLTI/PIONIER observations from 18/19 December 2020 (orbital phase 0.82).}\label{fig:interf_2020_2_CP}
\end{figure}

\begin{figure}[h]
\centering
\includegraphics[width=0.49\textwidth]{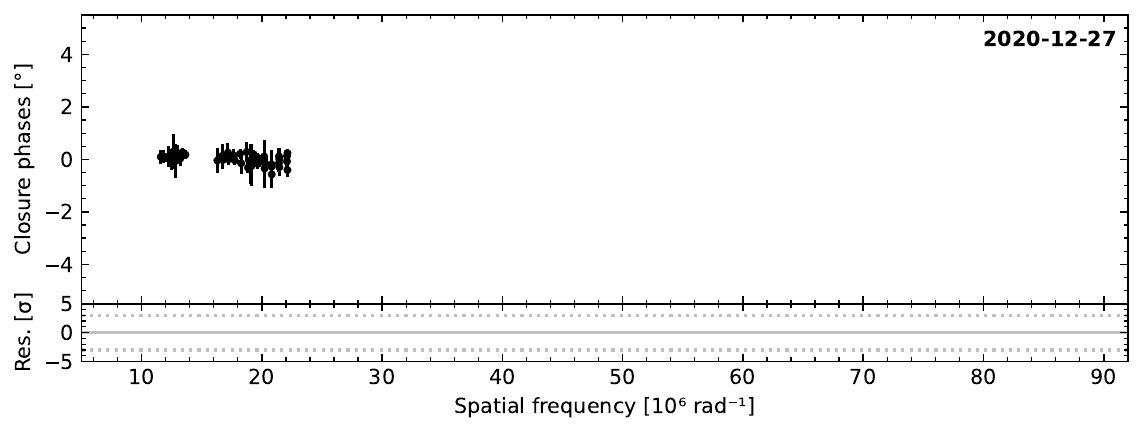}
\caption{Same as Fig. \ref{fig:interf_2010_CP}, but for VLTI/PIONIER observations from 27 December 2020 (orbital phase 0.91).}\label{fig:interf_2020_3_CP}
\end{figure}
\end{appendix}
\end{document}